\documentclass[prd,twocolumn,amssymb,noshowpacs,nofootinbib]{revtex4}
\usepackage{graphicx}
\usepackage{bm}

\begin{document}

\preprint{CMU-HEP-02-15}

\title{Radion Induced Baryogenesis}

\author{Matthew R.~Martin}
\email{mmartin@cmu.edu}
\affiliation{Department of Physics,
             Carnegie Mellon University,
             Pittsburgh, PA 15213}

\date{\today}

\begin{abstract}
Recent work in two brane Randall-Sundrum models has shown that the radion
can drive parametric amplification of particles on the TeV brane
leading to a stage of cosmology similar to preheating.  We consider
radion induced preheating as a mechanism for electroweak scale
baryogenesis and estimate the baryon to entropy ratio for some regions
of parameter space.  We also predict an upper bound for the radion mass,
which makes this baryogenesis mechanism testable.
Finally, we verify some assumptions with numerical calculations
of the 1+1 dimensional Abelian Higgs model.
\end{abstract}

\pacs{11.10.Kk,98.80.Cq,11.30.Fs}

\maketitle

\section{Introduction}
The high degree of symmetry between matter and anti-matter
in our theories has long seemed in contradiction with the
overwhelming presence of matter (baryons) around us.
Electroweak baryogenesis~\cite{reviews} provides
a testable explanation of the baryons within the context
of the standard models of particle physics and cosmology but
suffers from the two difficulties of
providing enough CP violation and providing enough departure from
equilibrium.  In the usual picture transitions between
vacua of the electroweak theory correspond, via an anomaly,
to changes in baryon plus lepton number.
If the electroweak phase transition in the early
universe is strongly first order then the presence of CP violation
will cause sphaleron transitions near bubble walls to favor formation
of matter over anti-matter.  While the minimal supersymmetric
standard model may provide the first order phase transition
and CP violation needed for this picture of baryogenesis,
a radically different picture is possible within the context
of the Randall-Sundrum model.

As with supersymmetry, the two-brane Randall-Sundrum model~\cite{rsa}
(RS) provides a framework for addressing the hierarchy problem.
The model consists of a slice of five dimensional anti-de Sitter 
space ($AdS_5$) truncated by two
3-branes with the standard model particles trapped on one of the
branes.  The gravitational warping between the branes generates
hierarchies of mass scales.  With the stabilization of the brane
separation provided by Goldberger and Wise~\cite{gw},
mass scales are naturally stabilized providing a solution
to the hierarchy problem along with an acceptable low-energy
phenomenology.  The separation between the branes
corresponds to a particle called the radion which
is expected to be the lightest new degree of freedom.

Initially, it might seem difficult to explain the baryon
asymmetry within the context of RS since RS physics
becomes important just above the electroweak scale.
For example, it is not possible to use existing extensions of
standard model physics to explain baryogenesis within an
effective theory that integrates out RS physics.
They are at the same scale.  We must, therefore, re-examine the
baryogenesis question within the context of the RS model.
Fortunately, the radion provides a new route for departure
from equilibrium; previous work has shown that homogeneous
oscillations of the radion lead to exponential growth of
scalar particles on the standard model brane~\cite{radion_preheating}.
The produced particles do not have a thermal distribution, but lie in a
small region of momentum space.  These are features common to
parametric amplification, or preheating~\cite{inflation_preheating},
as it occurs in the inflationary context.

We use parametric amplification of the standard model Higgs field
to drive the field over the potential barrier, allowing for
the creation of winding modes and, consequently vacuum
transitions.  This parametric amplification,
rather than bubbles from a first order phase transition, generates
the departure from equilibrium.  With an additional source of CP
violation, the decay of the Higgs windings will favor the
production of baryons over anti-baryons.  However, there are
constraints on the mass of the radion coming from the size of
windings which give rise to baryon number violating transitions.
Generically, if the radion is too massive, then the parametric
amplification gives windings which are too small for baryogenesis.  The
radion should not be much more massive than three times the mass of the
Higgs boson.  Furthermore, because the radion is a gravitational mode in the
full theory, the form of the radion-Higgs coupling is determined with only
one free parameter (we consider only minimal coupling of the Higgs).

The next sections review relevant aspects of the standard
model baryon violation and the RS model.  We then discuss radion
induced preheating and consequences for baryogenesis.
We explain some constraints on
the model and estimate the baryon asymmetry for a simple
case.  The final section shows numerical results which verify
some assumptions and help develop our intuition.

\section{Baryogenesis in the Standard Model}\label{sec:baryo}
As alluded to above, non-perturbative effects in the electroweak
sector of the standard model violate baryon number.  An anomaly
links the change in baryon and lepton number to the change in
Chern-Simons number:
\begin{equation}
\Delta B = \Delta L = n_f \Delta N_{CS},
\label{delta_Ncs}
\end{equation}
where $n_f$ is the number of families of matter fermions.  In the
absence of fermions the vacua of this theory are degenerate and
they are labeled by the winding number of the Higgs field.  In any
vacuum state the Chern-Simons number is equal to
the Higgs winding number.  The transition from
one vacuum state to another must therefore involve a change in
both the Higgs and gauge fields so that both the Higgs winding and
the Chern-Simons number remain equal.  See the reviews
in~\cite{reviews} for more details.  This baryon violating transition
between vacua passes through a field configuration known as the
sphaleron, which is a saddle point of the effective potential.
Henceforth, we refer to the sphaleron mediated transitions between
vacua as the sphaleron transition.

We can now understand that an initial configuration with the
Higgs winding differing from the Chern-Simons number by one unit must relax
to a vacuum in one of two ways: either the Higgs winding will change,
which does not violate baryon number; or the gauge field evolution
will change the Chern-Simons number, which does violate baryon number.
The method by which the relaxation occurs is dependent upon the
length scale of the winding.  In general a Higgs winding will collapse
and the time scale for the collapse is simply the length
scale~\cite{turok_zadrozny}.  However, the gauge fields will evolve
to cancel gradient energy in a Higgs winding with time scale loosely
set by the gauge boson mass.  So a sufficiently small winding will
collapse and the Higgs winding number will change, while a sufficiently large
winding will be canceled by the gauge field thus changing baryon
number.

Some critical length scale separates these two regimes.
If there is initially an equal density of windings and
anti-windings on large scales and no CP violation, then
there will be equal production of
baryons and anti-baryons.  According to the picture of Turok and
Zadrozny~\cite{turok_zadrozny}, CP violation will
affect the critical length differently
for windings and anti-windings.  An anti-winding near the
critical scale will preferentially unwind through Higgs field evolution, while a
winding at the same scale will unwind by gauge field evolution,
thus favoring production of baryons.
The critical length is about $3\,m_W^{-1} \sim (30\,GeV)^{-1}$ if
the gauge boson mass is given by the zero temperature vacuum.  The
realistic picture is more complicated~\cite{lrt} with the length
scale being a function of other parameters which characterize a
winding, but there should still be a range of critical length scales
separating Higgs field unwinding from baryon number changing.

In the high temperature, symmetry restored vacuum, simple dimensional
analysis gives the sphaleron length scale as
being the magnetic screening length, $(\alpha_W T)^{-1}$, which could
be as large as about $(3\,GeV)^{-1}$.  For the remainder of the paper
we work with the length scale $l_{sph} = (20\,GeV)^{-1}$.
We use this length scale to constrain the radion
mass since a massive radion will give higher frequency excitations
to the Higgs field leading to windings which are too small for baryogenesis.

At temperatures much below the electroweak scale, sphaleron transitions
are exponentially suppressed by the energy scale of the potential
barrier between electroweak vacua, $E_{sph} \sim 10\,TeV$:
\begin{equation}
\Gamma_{sph} \sim e^{(-E_{sph}/T)}.
\end{equation}
For temperatures near the electroweak scale and as low as
$150\,GeV$~\cite{preheat_lattice}, the electroweak symmetry is
restored and the sphaleron transitions are
unsuppressed.\footnote{Some authors~\cite{reviews} cite $100\,GeV$ as
the temperature above which sphaleron transitions are unsuppressed.
We choose to follow reference~\cite{preheat_lattice} since a larger
temperature is more conservative with respect to placing bounds on
the radion mass.}  Thermal fluctuations easily form windings of the
Higgs or gauge fields and on dimensional grounds
\begin{equation}
\Gamma_{sph} \sim (\alpha_W T)^4.
\end{equation}
It might seem then that a temperature above $150\,GeV$, along with CP
violation, is sufficient for the production of baryons.
However, the thermal average of baryon number is zero~\cite{reviews},
meaning that baryogenesis must happen during a departure from
thermal equilibrium.  Furthermore, if equilibrium is restored while
sphaleron transitions are unsuppressed, then any existing baryon
asymmetry will be destroyed.  This process is called thermal
washout, and it will constrain the total energy in our model as
explained in section~\ref{sec:preheat_baryo}.

\section{The Randall-Sundrum Scenario}
The background geometry for the two brane Randall-Sundrum model (RS)
is given by the metric
\begin{equation}
ds^2 = e^{-2kz}\eta_{\mu\nu}dx^{\mu}dx^{\nu} + dz^2,
\end{equation}
with branes truncating the space at $z=z_0$ and $z=z_1$.  This geometry
is a solution to Einstein's equations for negative bulk cosmological
constant and nonzero brane tensions which scale with $k$.  For
naturalness $k$ is taken to be near the fundamental Planck scale, $M_5$,
but somewhat smaller to ensure gravitational perturbativity.
The warping of the space red shifts mass scales on the brane at $z_1$
relative to the scale at $z_0$, allowing for the electroweak theory
trapped on the brane at $z_1$ to have an energy scale much less than
scales on the brane at $z_0$.  The brane at $z_1$ is therefore often
called the $TeV$ brane.  The separation between the branes corresponds to
a modulus known as the radion.  By fixing the radion in a natural way,
the electroweak hierarchy problem is solved.

The form for the radion which does not mix with the four dimensional
graviton~\cite{cgr} is
\begin{eqnarray}
ds^2 &=& e^{-2k \left( z+f(x)e^{2kz} \right) }
g_{\mu\nu}(x)dx^{\mu}dx^{\nu} \nonumber \\
&& + \left( 1+2kf(x)e^{2kz} \right) ^2dz^2,
\end{eqnarray}
where $f(x)$ is the radion.  Goldberger and Wise~\cite{gw} found
that a scalar field in the bulk which has nonzero vacuum
expectation values on the branes can naturally determine the brane
separation.  A competition between the kinetic and potential terms
in the five dimensional action for the scalar gives a potential to the radion
in the four dimensional effective theory.  In other words, after integrating
out the bulk degrees of freedom, an effective theory for the radion
remains~\cite{radion_preheating}:
\begin{equation}
\mathcal{L}_F = -\frac{1}{2}\partial_{\mu}F(x)\partial^{\mu}F(x)
-\frac{1}{2}m_F^2F(x)^2 + \mathcal{O}(F^3).
\end{equation}
We have rescaled $f(x)$ to have the properly normalized kinetic term,
\begin{equation}
F(x) = \sqrt{12 k M_5^3}e^{k(z_1-z_0)}f(x)
\end{equation}
and the value for the radion mass, $m_F$, depends on the details of
the bulk scalar.  The scale of physics on the $TeV$ brane is set by the
parameter $\Lambda = \sqrt{3M_5^3/k}e^{-k(z_1-z_0)}$.  We will only consider
small oscillations of the radion, $F(x) \ll \Lambda$, to avoid
sensitivity to the higher order terms in the radion effective action.

Now consider a scalar field on the $TeV$ brane.  We will see how the
warping red shifts mass scales and derive the coupling of the radion to
the scalar needed for the next section.  We set $z_0 = 0$ for simplicity.
The action on the brane is
\begin{equation}
S = -\frac{1}{2} \int d^4x \sqrt{-\tilde{g}} \left[
\tilde{g}^{\mu\nu}\partial_{\mu}\phi\partial_{\nu}\phi
+ m^2\phi^2 \right],
\end{equation}
with the induced metric on the brane given by
\begin{equation}
ds^2 = \tilde{g}_{\mu\nu}(x)dx^{\mu}dx^{\nu} =
e^{-2k \left( z_1+f(x)e^{2kz_1} \right) }\eta_{\mu\nu}dx^{\mu}dx^{\nu}.
\end{equation}
By rescaling $\phi(x) \rightarrow e^{kz_1}\phi(x)$ to give the
kinetic term canonical normalization, the action takes the form:
\begin{eqnarray}
S &=& - \frac{1}{2} \int d^4x\, \Bigl[
e^{-2kf(x)e^{2kz_1}} \eta^{\mu\nu}
\partial_\mu\phi\partial_\nu\phi \nonumber \\
& &\qquad\qquad\quad + e^{-4kf(x)e^{2kz_1}}e^{-2kz_1}
m^2\phi^2 \Bigr].
\end{eqnarray}
Notice that the scalar field mass is suppressed by a ``warp factor'',
$e^{-kz_1}$.  We define the effective scalar mass $m_\phi = e^{-kz_1}m$.
In terms of the correctly normalized radion, the action is:
\begin{eqnarray}
S &=& -\frac{1}{2} \int d^4x\, \Bigl[
e^{-F(x)/\Lambda} \eta^{\mu\nu}
\partial_\mu\phi\partial_\nu\phi \nonumber \\
& &\qquad\qquad\quad + e^{-2F(x)/\Lambda}
m_\phi^2\phi^2 \Bigr] .
\label{branescalar}
\end{eqnarray}
This is the action used to study parametric amplification of
scalars on the $TeV$ brane~\cite{radion_preheating}.  Notice that
the universality of gravitational couplings limits the
radion-scalar coupling to this form.  In the next
section we review this parametric amplification and show how
baryogenesis occurs.

\section{Preheating and Baryogenesis}\label{sec:preheat_baryo}
As we discussed in previous work~\cite{radion_preheating}, there may have
been a time in the early universe when most of the energy density
was tied up in coherent homogeneous oscillations of the radion.  These
oscillations might be expected for several reasons.  First, an inflation-like
mechanism is needed to solve the horizon problem within the RS
context, and this mechanism almost certainly will involve physics of the
extra dimension.  In a minimal model the only way to reheat the
$TeV$ brane is through the stabilization mechanism, which would
involve oscillations of the stabilizing field and of the radion.
We take the potential of the stabilizing field to be stiff and study
oscillations of the radion.

Alternatively, at high enough temperatures
the space-time is unstable to formation of a bulk horizon, which cuts
off the space and removes the $TeV$ brane.  As the universe cools, a
phase transition occurs~\cite{rattazzi}, generating the $TeV$ brane
generically shifted from its equilibrium position.  The radion will oscillate
coherently as the brane settles in to the equilibrium position.
In this scenario we would not expect a cold $TeV$ brane, although more
details of this phase transition still need to be understood.

In a third picture, the stabilization mechanism may provide a false
vacuum for the radion with the $TeV$ brane at infinity\cite{cline}.
Early universe dynamics could set the brane in this false
vacuum with a first order phase transition bringing the brane
to the global minimum.  Bubbles of the brane in the
true vacuum would not be expected to lie exactly at the bottom
of the potential, but oscillate around the minimum.
In any of these three scenarios, higher energy physics leads
to radion oscillations.

\begin{figure}
\includegraphics{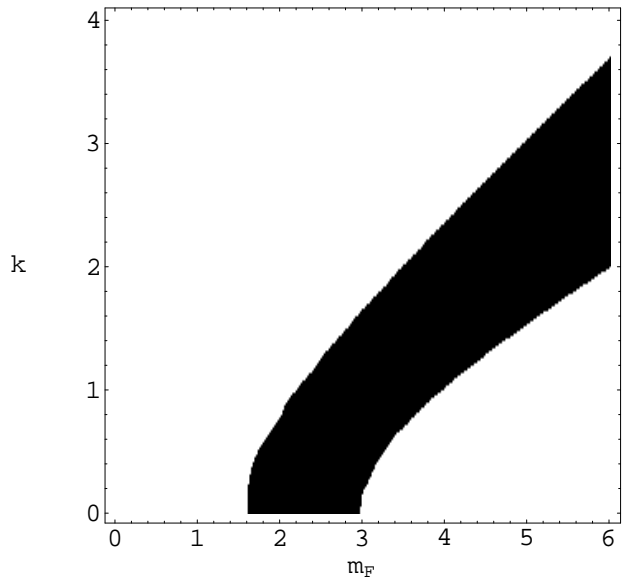}
\caption{The instability band for the massive scalar
when $F_0=0.5$.  The dark region shows where there is
exponential growth of scalar particles.  Both $m_F$ and $k=|\vec
k|$ are expressed in units of the scalar mass.\label{higgs_band}}
\end{figure}

We take the radion to be a background of the form 
\begin{equation}
F(t) = F_0 \Lambda \cos(m_F t),
\label{radion_osc}
\end{equation}
and decompose the scalar field in Fourier modes:
\begin{equation}
\phi(t,\vec x) = \int {d^3k\over (2\pi)^{3/2}}\, \left[ a_{\vec k}
\phi_k(t) e^{-i\vec k\cdot\vec x} + a_{\vec k}^\dagger \phi_k^*(t)
e^{i\vec k\cdot\vec x} \right] .
\end{equation}
We then calculate the equation of motion for the scalar modes from
the action~(\ref{branescalar}):
\begin{equation}
{d^2\phi_k \over dt^2} + F_0 m_F \sin(m_F t) {d\phi_k
\over dt} + {\omega}^2_k (t) \phi_k = 0,
\label{phi_k_eom}
\end{equation}
where ${\omega^2_k (t)}\equiv \left[ |\vec k|^2 + m_\phi^2 e^{-F_0\cos(m_F t)}
\right]$.  We start with a vacuum state for all modes,
\begin{equation}
{d\phi_k \over dt}(0) = -i\omega_k(0)\phi_k(0) ,
\quad \phi_k(0) = \frac{e^{F_0/2}}{\sqrt{2\omega_k(0)}},
\label{initial_conditions}
\end{equation}
and find exponential growth of some modes.  For the specific case
$F_0=0.5$, the modes unstable to growth are shown in the dark region
of figure~\ref{higgs_band}.  This is parametric resonance, or preheating.

It should be emphasized that the spectrum of particles produced
through parametric amplification is non-thermal
and the time scale for growth of particles is given by the radion
mass.  To show this we calculate the expectation value of the number
density operator, $\langle N_k(t) \rangle$ (see the appendix for details).
In figure~\ref{nks} we plot $\ln \langle N_k(t) \rangle$.
The modes in momentum space which grow are those
expected by comparing with figure~\ref{higgs_band}.  We expect
that when backreaction of the scalar on the radion is taken into
account, the scalar will drain energy off the radion and damp the
amplitude of radion oscillations.

\begin{figure}
\includegraphics{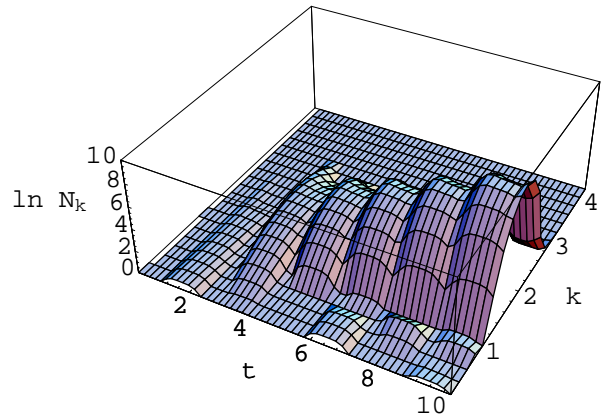}
\caption{Log of the number density operator, $\ln N_k(t)$, for
$m_F=4m_\phi$ and $F_0=0.5$.  Note that for
momenta within the instability band of figure~\ref{higgs_band}, the number of
particles produced grows exponentially.  All units are appropriate
powers of the scalar mass, $m_\phi$.
\label{nks}}
\end{figure}

We now want to think about placing the Standard Model on the
$TeV$ brane.  Again assuming homogeneity and an initial state
near the zero temperature vacuum, we would expect Higgs particle production as
explained above, where $m_\phi$ is now the Higgs mass.  Of
course, the action for the Higgs field should involve the self-coupling
and the coupling to gauge fields.  However, leaving these terms out should 
be a good approximation at the beginning when only
the minimum of the potential is being sampled by the Higgs vacuum
expectation value.
The choice of initial conditions is generalized in the
numerical work.  We discuss production of fermions and
gauge fields below.

The essence of the idea is that the
preheating of the Higgs field will produce configurations with
winding, in a manner analogous to the production of topological
defects following inflationary preheating.  These windings will then decay.
Some decays will violate baryon number, some will not, and the
CP violation will favor production of a small number of net
baryons.  A similar idea of inflationary preheating leading
to baryogenesis has been
proposed by others~\cite{inflate_preheat_baryo1,inflate_preheat_baryo2}.
However, radion preheating occurs naturally at the proper
energy scale, the electroweak scale, and has little freedom
in parameter space due to the universality of radion couplings.
These concepts are significant: there is no fine tuning needed, as
in the inflationary case, to make radion preheating happen at the correct
energy scale; and our model will be tested and may be excluded at
the next generation of colliders.

\subsection{Constraints}
Before estimating the amount of baryogenesis, we would like to discuss
some constraints on the radion.  First, the final reheat temperature should be
sufficiently below the electroweak phase transition to avoid thermal
washout of the baryon asymmetry.  If, once thermal equilibrium is
reached, we require
\begin{equation}
T_{reheat} \lesssim 150 \, GeV,
\label{T_reheat}
\end{equation}
then we must constrain the total energy density:
\begin{equation}
\rho = \frac{\pi^2 g_*}{30} T^4 \lesssim (350 \, GeV)^4.
\label{energy_temp}
\end{equation}
Because the expansion rate of the universe,
$H \sim \frac{\sqrt{\rho}}{M_{Pl}} \sim 10^{-15}GeV$,
is much less than other relevant scales in the problem,
we may ignore the expansion.
Then energy conservation restricts the initial radion configuration:
\begin{equation}
\rho_F = \frac{1}{2}F_0^2 \Lambda^2 m_F^2 \lesssim (350 \, GeV)^4.
\label{radion_energy_density}
\end{equation}

\begin{figure}
\includegraphics{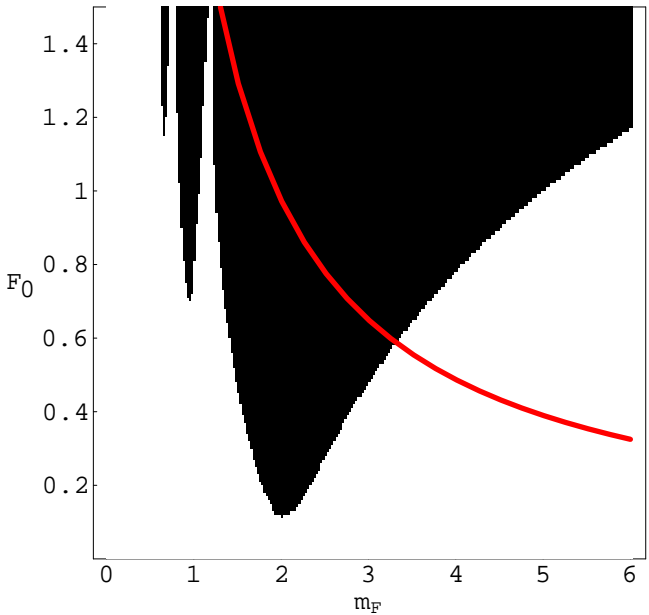}
\caption{The instability band for the massive scalar
when $k=0.16\,m_{\phi}$.  This corresponds to $k=l_{sph}^{-1}$ for
a Higgs mass of $125\,GeV$.  The dark region shows where there is
exponential growth of particles.  The radion mass
is expressed in units of the Higgs mass.
Above the line the system contains enough energy
to washout baryons after thermalization.\label{higgs_band_k}}
\end{figure}

Another constraint comes from the location of the instability
band.  As explained in
section~\ref{sec:baryo}, the length scale for sphaleron
configurations, $l_{sph}$, is about $(20 \, GeV)^{-1}$.
The instability band therefore needs to encompass this value
of momentum in order to produce sphaleron configurations.
For the case considered in figure~\ref{higgs_band}, this criteria is only
satisfied for a radion mass of $1.5 \lesssim m_F/m_{\phi} \lesssim 3$.
More generally, we can understand this constraint
by plotting the instability band
as a function of $F_0$ and $m_F$ for fixed $k = l_{sph}^{-1}$ as in
figure~\ref{higgs_band_k}.

The region above the line is excluded by the energy
density constraint.  This assumes a
Higgs mass of $125\,GeV$, to set a scale, and a
fairly low value for the cutoff, $\Lambda = 750\,GeV$.  These
values are used in the remainder of the paper.  Equation
(\ref{radion_energy_density}) shows that this line moves down
(excluding more of the preheating band) as
the lower limit of the cutoff increases or if the Higgs mass
becomes heavier.  Already we can see that if the radion mass is much
above three times the Higgs mass, then there can be no preheating
without thermal washout of baryons.  For a light
radion we need large values of $F_0$ for preheating.  To trust
a calculation in this regime requires knowledge of the complete
radion potential.  This involves details of the stabilizing
mechanism.

While $\mathcal{O}(F^3)$ and higher terms in the radion action 
will alter the form of radion oscillations and make a
precise large $F_0$ calculation of the baryon to entropy ratio difficult,
in practice, these terms do not prohibit parametric
amplification of the Higgs field.  A simple numerical calculation
similar to that shown in figure~\ref{nks} demonstrates that
arbitrary $\mathcal{O}(\Lambda)$ and $\mathcal{O}(1)$ coefficients for
$F^3$ and $F^4$ terms respectively preserve exponential
growth of the Higgs fluctuations.  The significant concern for the
purposes of this paper is that higher terms would modify the
location of the instability band without a comparable modification
of the energy density constraint, thus altering the
upper limit on the radion mass.  While we can't rule out this
possibility, it does not seem to be a generic feature of these
operators.

\subsection{Evolution}
We may estimate when sufficient preheating has occurred to
form Higgs windings by calculating the two point correlation
function of the Higgs fluctuations.  We work in a toy
model with no gauge fields.  Shifting the Higgs field,
\begin{equation}
\phi = \varphi + \chi, \quad \varphi \equiv \langle\phi\rangle,
\end{equation}
the equation of motion for the vacuum expectation value (VEV)
of the Higgs in the
Hartree approximation~\cite{rich_dan} becomes:
\begin{equation}
\Box \varphi + \lambda \left( \varphi^2-v^2+3 \langle \chi^2 \rangle
\right)\varphi = 0.
\end{equation}
In the electroweak theory windings will be able to form when
the Higgs VEV crosses the potential barrier.  In this toy theory,
$\varphi$ crosses the barrier when this two point function is
of the order of the zero temperature vacuum expectation value:
\begin{equation}
\langle \chi^2 \rangle \sim v^2/3.
\label{winding_condition}
\end{equation}
It is not clear that the system will necessarily behave as if
the symmetry of the vacuum for $\varphi$ has been restored, but we
do not need to concern ourselves with this issue.  We only need
the fluctuations, $\langle \chi^2 \rangle$, to kick the field across
the potential barrier for winding formation.

\begin{figure}
\includegraphics{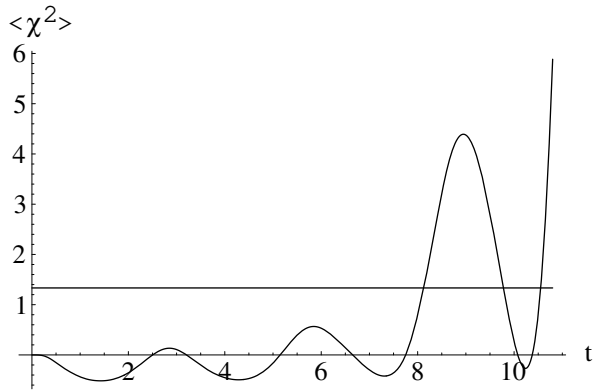}
\caption{The two point function for the Higgs fluctuations
with $m_F = 2.25\,m_{\phi}$ and $F_0=0.5$ with units in powers
of the scalar mass.  The horizontal line
is $v^2/3$ when $m_{\phi} = 125\,GeV$.}
\label{condensate}
\end{figure}

In figure~\ref{condensate} we plot $\langle \chi^2 \rangle(t) - 
\langle \chi^2 \rangle(0)$ where
\begin{equation}
\langle \chi^2 \rangle (t) = \int_0^{\Lambda} \frac{d^3k}{(2\pi)^3}
|\chi_k(t) |^2.
\end{equation}
The equation of motion for the Fourier modes, $\chi_k$, is given by
equation~(\ref{phi_k_eom}) where $\phi_k$ is replaced by $\chi_k$ and
$m_\phi$ is now the Higgs mass.  We start again with vacuum initial
conditions as in equation~(\ref{initial_conditions}).
From the plot it is clear that the winding formation condition,
equation (\ref{winding_condition}), is satisfied after a few radion
oscillations.  Of course, at this point the approximation of
linearizing the equations has broken down.
We are simply making the point here that windings should form
on a time scale set by the radion mass.
Because $\langle\chi^2\rangle$ is cutoff dependent
even in the initial vacuum state, we plot the
change in this quantity.

There are several possible times scales relevant to this problem.
First, the Hubble scale, is often important in electroweak baryogenesis
since the Hubble expansion cools the universe and causes the
sphaleron transitions to freeze out.  Here that role is played by
the thermalization time scale and the Hubble parameter is irrelevant.
The sphaleron transitions freeze out as the energy in the Higgs field is
redistributed to all the other degrees of
freedom in the standard model.  The total decay rate of the Higgs boson is near
$10^{-2}\,GeV$~\cite{carena_haber} so we assume this determines the
thermalization time scale.  We take the sphaleron time scale to be the
length scale, $l_{sph}$, which is much shorter than the thermalization
scale.  Finally, the fourth time scale is that of preheating.
As we have seen above, the preheating scale is given by the radion mass.

This picture is not significantly altered by
the presence of matter fermions or gauge fields.
Pauli blocking prevents the exponential growth of fermions,
and the decay time scale of the Higgs boson is much longer than the preheating
scale.  Fermions can safely be ignored.
The radion also couples to the gauge fields, but
not to the kinetic term.  Without the derivative
coupling that was present between the radion and the Higgs in
equation~(\ref{phi_k_eom}),
there is much less parametric amplification.
The full sphaleron transition naturally involves gauge fields, but here
we focus on generation of the Higgs windings only and leave
more detailed study for future work.

\subsection{An Estimate}
For generic parameters inside the preheating band of figure~\ref{higgs_band_k},
the radion will continue to produce windings of the appropriate
length scale for much longer than the sphaleron time scale.  This
winding production time will depend on the total radion energy and on
the thermalization time scale for the system.  For example, in
figure~\ref{condensate} only about $1/10$ of the energy of the radion
has gone in to the Higgs field by the time windings form.
We expect baryon production to be quite efficient here, since sphaleron
transitions may take place throughout the entire brane during preheating.
To calculate the baryon asymmetry, though, will require a deeper understanding
of the sphaleron transition rate in this far from equilibrium
state.  It will also require the full $SU(2)$ Higgs and gauge fields,
a better understanding of the thermalization time scale,
and, of course, a source of CP violation.
There has been work on this calculation within the context of
baryogenesis following inflation.  Some efforts have applied
ideas from thermal sphaleron transitions~\cite{inflate_preheat_baryo2},
while others have applied lattice calculations~\cite{preheat_lattice,
preheat_lattice2}.  These steps are important, but difficulties may
remain~\cite{lattice_problems}.

\begin{figure}
\includegraphics{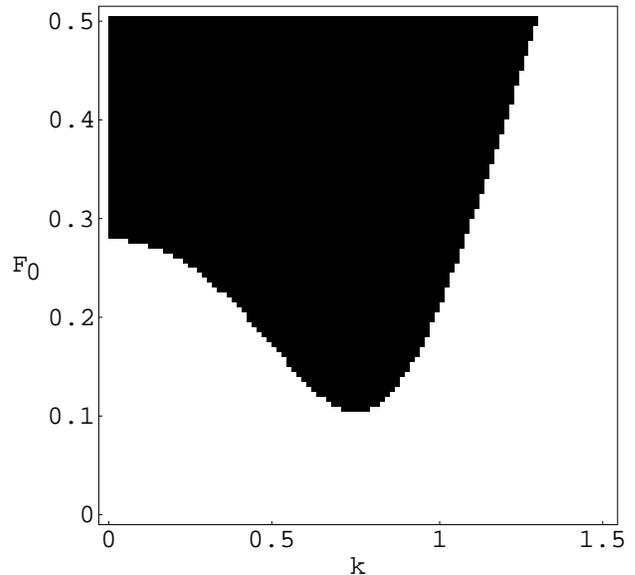}
\caption{The $m_F = 2.5\,m_{\phi}$ plane of the instability band.
The momentum, $k$, is in units of Higgs mass.
\label{higgs_band_w}}
\end{figure}

Nevertheless, for some specific choices of $F_0$ and $m_F$, it may be
possible to estimate the winding density of the Higgs field and acquire
insight on the baryon asymmetry in a manner which is
independent of a new CP violating mechanism.  If the radion preheating
generates windings and shuts off on a time scale shorter than
the sphaleron time, then the Higgs winding density can
be estimated via the Kibble mechanism.  To see how this might
happen, we plot the instability band
as a function of $F_0$ and momentum in figure~\ref{higgs_band_w}
for radion mass $m_F = 2.5\,m_{\phi}$.  Considering the backreaction
of the Higgs field on the radion, we expect a smaller $F_0$ to be relevant
later as the radion loses energy to the Higgs field.

So, if the system starts near $F_0 = 0.3$, then sphaleron sized
windings may occur for a time $\sim l_{sph}$.  As $F_0$ decreases, much of
the remaining energy of the radion will be dumped to higher momentum
modes centered around $k=0.75\,m_{\phi}$.  Then, much as happens
in defect formation through inflationary preheating, we
expect that there will be order one winding per $l_{sph}^3$
region of space.  In this process the quantum fluctuations
are being amplified, and unless there is reason to expect
long range correlations, then causally disconnected
regions should wind in different directions.

We may now make our estimate.  We calculate the entropy present:
\begin{equation}
s = \frac{2\pi^2}{45}g_{*s}T^3,
\end{equation}
where the temperature after thermalization is given by the energy density
for $F_0 = 0.3$, and $m_F = 2.5m_{\phi}$, along with
equation~(\ref{energy_temp}): $T \approx 0.75\,m_{\phi}$.  We parameterize the
CP violation such that the baryon density is the product of the winding
density and $\epsilon_{cp}$:
\begin{equation}
B = n_w \epsilon_{cp}.
\end{equation}
Then the baryon to entropy ratio becomes
\begin{equation}
\eta = \frac{B}{s} \approx \frac{l_{sph}^{-3} \epsilon_{cp}}{s}
\approx 10^{-4} \epsilon_{cp}
\end{equation}
The measured value is $\eta \approx 10^{-10}$.
Note that the strong dependence of the winding density to the sphaleron time
scale makes this estimate imprecise.

It is common in four dimensional scenarios to postulate that a
nonrenormalizable CP violating term arises from a higher energy
theory.  The same may be possible here, but any higher energy theories
need to live in the bulk.  It would be interesting to investigate
the details of bulk CP violation and the consequences for
baryogenesis on the brane.  It may also be possible to add the CP
violation on the brane at electroweak energies, as in a two Higgs
doublet model.

\section{A Numerical Model}
To see some details of this process borne out more
explicitly, we now turn to a numerical implementation of
the 1+1 dimensional Abelian Higgs model.  This model has
features in common with the baryon violation in the standard
model, as we will see below, and has therefore often been used in the past
to study baryogenesis~\cite{reviews}.  The Abelian Higgs model allows us
to go beyond the linearized equations and to include backreaction 
on the radion without the technical difficulties of
solving a non-Abelian model numerically.  We can study the formation of
winding modes and confirm our intuition that the length scale
of the windings is given by the location of the instability
band.

The action for our toy theory is
\begin{eqnarray}
S &=& \int d^2x \Bigl(-\left(\mathcal{D}_{\mu}\phi\right)^{\dagger}
\mathcal{D}^{\mu}\phi -\textstyle{1\over 4}F_{\mu\nu}F^{\mu\nu}
\nonumber \\
&& \qquad\qquad -\textstyle{\lambda\over 4}\left(\phi^{\dagger}\phi
-v^2\right)^2 \Bigr) .
\end{eqnarray}
In the vacuum the scalar sits at the bottom of the potential well
\begin{equation}
\phi = ve^{i \theta(t,x)}
\end{equation}
and $A^{\mu}$ is pure gauge
\begin{equation}
A_{\mu} = \textstyle{1\over e} \partial_{\mu}\theta(t,x) .
\end{equation}
This theory has a winding number which labels the vacua
\begin{equation}
N_H(t) = \textstyle{1\over{2\pi}} \int dx \, \partial_x \theta,
\end{equation}
and transitions between vacua require $\phi$ to pass through zero.
In addition, the Chern-Simons number is proportional to the
Higgs winding number in the vacuum state:
\begin{equation}
N_{CS} = \int dx \, A_x = \textstyle{2\pi\over e} N_H .
\end{equation}
Thus the vacuum structure of this theory resembles that of the
electroweak sector of the standard model.  With the addition of
fermions there is a conserved chiral current, which is anomalously
broken in a way that is related to the change in Chern-Simons
number~\cite{peskin}, $\Delta (N_R - N_L) = \Delta N_{CS}$.

Because the radion, which stabilizes a slice of $AdS_3$, does
not couple as the 5 dimensional radion,
we introduce a ``radion'' designed to mimic the 3+1 dimensional preheating.
As mentioned above, we are primarily interested in the early
time behavior and specifically the formation of the
Higgs windings modes.  Therefore, we restrict our attention to
the radion-Higgs sector:
\begin{eqnarray}
S &=& \int d^2x \Bigl( -\textstyle{1\over 2} e^{-F/\Lambda} \eta^{\mu\nu}
\partial_\mu\phi^{\dagger}\partial_\nu\phi \nonumber \\
& &\qquad\qquad - {\textstyle{\lambda\over 4}} e^{-2F/\Lambda}
\left(\phi^{\dagger}\phi-v^2\right)^2 \Bigr) . \quad
\end{eqnarray}

We start the Higgs field in a vacuum state as before.
We then briefly evolve the linearized quantum equations~(\ref{phi_k_eom}).
After particle production has begun, but before
equation~(\ref{winding_condition}) is satisfied,
we Fourier transform to configuration
space and evolve the full non-linear classical equations of
motion.  In other words, the quantum field
with a few particles is serving as
the initial configuration of the classical field for our
classical evolution.\footnote{After our work was completed,
reference~\cite{smit_tranberg} appeared using the same method of
setting initial conditions for the classical field.}
We checked that results are not sensitive
to the time when the transition from quantum to classical evolution
occurs.  Because of the exponential growth in the preheating
band, results are also relatively insensitive to random changes in the
initial conditions, equation~(\ref{initial_conditions}), by factors
of 4 or more.  With such large initial fluctuations, there are more
windings at early times, but after several oscillations of the radion
there is not a significant difference.  However, an initially hot brane,
or a brane with any non-trivial initial energy density, does make a change
to the reheat constraint, equation~(\ref{radion_energy_density}) since
it is the total energy density which is constrained by
equation~(\ref{energy_temp}).  So initial energy density on the brane
will move the line in figure~\ref{higgs_band_k} down, thus excluding
more of the preheating band.

\begin{figure}
\includegraphics{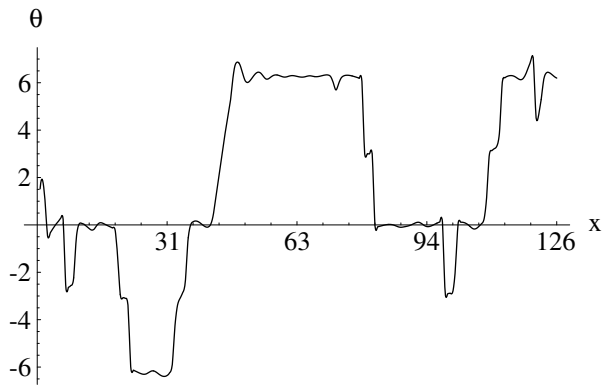}
\caption{Phase of Higgs field shortly after windings begin to form.
The length, $x$, is in units of inverse Higgs mass.
\label{phase}}
\end{figure}

\begin{figure}
\includegraphics{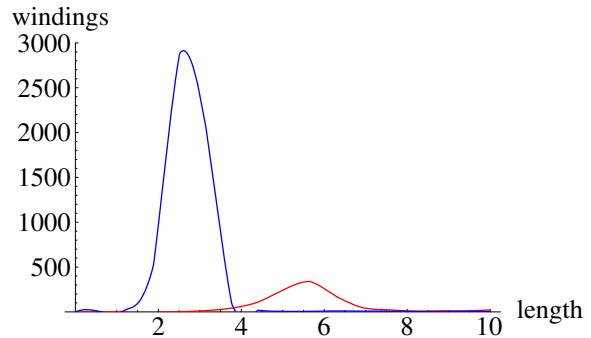}
\caption{Number and length scale of windings a few time units after
winding formation in units of $m_{\phi}^{-1}$.  Both use $F_0= 0.5$,
but have different radion mass: $m_F = 4.0\,m_{\phi}$ on the left
and $m_F = 2.0\,m_{\phi}$ on the right.
The length scale relevant for baryogenesis, $l_{sph}$, is in the range
of $6\,m_{\phi}^{-1}$.  The total length for the simulation
was $10^5\,m_{\phi}^{-1}$.
\label{winding_density}}
\end{figure}

In figure~\ref{phase} we
show the phase of the Higgs field after windings have begun to form.
We want to know the length scales
of the windings.  At early times, as in figure~\ref{phase},
windings are easy to identify since the fluctuations in phase
have not grown too large.  We define the length of winding as
being the minimum distance between two points separated in
phase by $2\pi$.  In
figure~\ref{winding_density} we show the number
of windings as a function of length scale for two cases
with $F_0 = 0.5$, as in figure~\ref{higgs_band}.  The curve
peaked at smaller scales corresponds to the more massive
radion, which has a preheating band at larger values of momentum.
This is what we expect.  For the case with longer length windings,
$m_F = 2.0\,m_{\phi}$, there are about 1200 windings.  This
corresponds to about one winding in every 10 causal volumes.  The
other scenario, $m_F = 4.0\,m_{\phi}$, preheats more efficiently,
in part because the radion oscillates more quickly leading to
about one winding in every 2 causal volumes.
In both cases the winding density is increasing quickly with time.

It would be interesting to use numerical methods on the classical
3+1 dimensional $SU(2)$ theory to track the Chern-Simons and Higgs
winding numbers.  The dependence of these quantities on the radion
mass should unambiguously confirm or refute the model.  In addition,
this calculation would provide a more concrete estimate of the
range of baryon to entropy ratios possible.

\section{Conclusion}
We have shown that baryogenesis can occur in the two-brane
Randall-Sundrum model with radion preheating
providing the departure from equilibrium.  Another
source of CP violation is needed, and as mentioned, investigations
of bulk CP violation may prove interesting.
One of the successes of this model is that
preheating occurs naturally at the energy scale for sphaleron
transitions to generate the asymmetry, but not wash it out.

Furthermore, the existence of a baryon asymmetry in our universe
implies an upper bound on the radion mass.
For a sufficiently large radion mass, the
preheating band does not extend down to sphaleron length scales,
unless the radion energy exceeds the thermal washout limit.  We
conclude that the radion mass should not be much larger than
three times the Higgs mass, if the baryon asymmetry is to arise
from this model.

\appendix*
\section{Number Density Operator}
We calculate here the expectation value of the number density
operator, $\langle 0| \hat{a}^{\dagger}_k(t) \hat{a}_k(t) |0 \rangle$,
for a massive
scalar field coupled to the radion, i.e. for action~(\ref{branescalar}).
First, let us change the notation slightly to ease this calculation.  We
decompose the scalar:
\begin{equation}
\phi(t,\vec{x}) = \int \frac{d^3k}{(2\pi)^{3/2}} \hat{\phi}_k(t)
e^{-i\vec{k}\cdot\vec{x}},
\end{equation}
where
\begin{equation}
\hat{\phi}_k(t) = U_k(t)\hat{a}_k + U^*_{-k}(t)\hat{a}^{\dagger}_{-k}.
\end{equation}
We decompose the conjugate momentum, $\pi(t,\vec x)$, in the analogous
manner and find
\begin{equation}
\hat{\pi}_k(t) = e^{-F(t)/\Lambda}\left( \dot{U}_k(t)\hat{a}_k
+ \dot{U}^*_{-k}(t)\hat{a}^{\dagger}_{-k} \right).
\end{equation}
We have assumed that the radion is independent of the
spatial coordinate in anticipation of homogeneous oscillations.

This decomposition is in terms of operators which 
annihilate the initial vacuum, $\hat{a}_k |0\rangle = 0$.  We seek
an expression for time dependent operators, $\hat{a}_k(t)$, in terms
of the initial time operators so that we may calculate the number of
particles relative to the initial vacuum.  The time dependent (Heisenberg
picture) operators must satisfy the equation of motion
\begin{equation}
i \frac{d}{dt} \hat{a}_k(t) = \left[ \hat{a}_k(t), H(t) \right],
\end{equation}
where $H(t)$ is the Hamiltonian.
The Hamiltonian can be written in terms of $\hat{\phi}_k(t)$
and $\hat{\pi}_k(t)$:
\begin{eqnarray}
H(t) = && \frac{1}{2} \int \frac{d^3k}{(2\pi)^{3/2}} \Big[
e^{F(t)/\Lambda}\hat{\pi}_k(t)\hat{\pi}_{-k}(t) \nonumber \\
&& \qquad + e^{-F(t)/\Lambda}\omega_k^2(t)
\hat{\phi}_k(t)\hat{\phi}_{-k}(t) \Big].
\end{eqnarray}
As in the main text, $\omega_k^2(t) = k^2 + m_{\phi}^2e^{-F(t)/\Lambda}$.
We may look for a solution by making the ansatz:
\begin{equation}
\hat{a}_k(t) = g(t)\hat{\pi}_k(t) + h(t)\hat{\phi}_k(t).
\label{ansatz}
\end{equation}
This guess is motivated by the form of a Bogolubov transformation
and the knowledge that $\hat{\pi}_k(t)$ and $\hat{\phi}_k(t)$
contain all of the degrees of freedom of the system.

From the initial conditions, $\hat{a}_k(t\!=\!0) = \hat{a}_k$, and
\begin{equation}
{d U_k \over dt}(0) = -i\omega_k(0)U_k(0) ,
\quad U_k(0) = \frac{e^{F(0)/2/\Lambda}}{\sqrt{2\omega_k(0)}},
\end{equation}
as in equation~(\ref{initial_conditions}), we find the
initial values:
\begin{equation}
g(0) = iU^*_k(0) \quad h(0) = -ie^{-F(0)/\Lambda}\dot{U}^*_k(0).
\end{equation}
Finally, using the standard equal time commutation relations for the field,
$[\hat{\phi}_{\vec{k}}(t),\hat{\pi}_{\vec{p}}(t)]
 = i(2\pi)^3\delta^3(\vec{k}+\vec{p})$ we may evaluate the
equation of motion for $\hat{a}_k(t)$.  The result is simple:
\begin{equation}
{d \over dt}g(t) = 0 = {d \over dt}h(t).
\end{equation}
We now use the form for the oscillating radion given in the text in
equation~(\ref{radion_osc}) and evaluate the number operator from
the ansatz~(\ref{ansatz}):
\begin{eqnarray}
&& \langle N_k(t) \rangle = \langle 0| \hat{a}_k^{\dagger}(t)
\hat{a}_k(t) |0 \rangle = \nonumber \\ &&\quad
{e^{-F_0}\over 2\omega_k(0) } \left[
e^{2F_0\left[ 1-\cos(m_F t)\right]} |\dot{U}_k(t)|^2
+ \omega_k^2(0) |U_k(t)|^2 \right] \ \ \nonumber \\
&&\quad
- {ie^{-F_0\cos(m_F t)}\over 2} \left[ U_k^*(t)
\dot{U}_k(t) - \dot{U}_k^*(t) U_k(t)
\right] . \ \label{Nfork}
\end{eqnarray}

\begin{acknowledgments}
I would like to thank Rich Holman for valuable discussion and Mark
Trodden for useful comments on the manuscript.
This work was supported by DOE grant DE-FG03-91-ER40682.
\end{acknowledgments}

\end{document}